\documentclass[12pt, aps,prl,showpacs, preprint, groupedaddress]{revtex4-1}
\usepackage{epsfig}
\usepackage{latexsym}
\usepackage{graphicx}  % needed for figures
\usepackage{dcolumn}   % needed for some tables
\usepackage{bm}        % for math
\usepackage{amssymb}   % for math
\usepackage{indentfirst}
\usepackage{subfigure}
\newcommand{\EQ}{\begin{equation}}
\newcommand{\EN}{\end{equation}}
\newcommand{\EQA}{\begin{eqnarray}}
\newcommand{\ENA}{\end{eqnarray}}

\begin{document}

%\preprint{}

%\leftline{To be submitted to Phys. Rev. Lett. or a letter for Phys. Fluids}
\title{Clustering and relative velocities of heavy particles under gravitational settling in isotropic turbulent flows}


\author{Guodong Jin}
%\email[]{gdjin@lnm.imech.ac.cn}
\author {Guowei He}
%\email[]{hgw@lnm.imech.ac.cn}
%\thanks{}
\affiliation{LNM, Institute of Mechanics, Chinese Academy of Sciences, Beijing 100190, P. R. China}


\date{\today}
\begin{abstract}
Spatial clustering and intermittency in the relative velocity of heavy particles of the same size settling in turbulent flows can be strongly affected by gravity. We present a model for the timescale of the fluid velocity gradient seen by particle pairs and propose an effective Kubo number based on this timescale to explain the mechanism of gravity-enhanced clustering. We explore the mechanisms of the gravity-induced reduction or enhancement of the intermittency in the particle radial relative velocity (RRV) at different Stokes numbers based on backward-in-time relative dispersion and preferential sampling of the fluid field. These effects of gravity on clustering and the RRV must be parameterized in the geometric collision kernel.

\end{abstract}

% insert suggested PACS numbers in braces on next l
\pacs{47.27.Gs, 42.68.Ge, 47.55.Kf, 92.60.Vb}
%\keywords{}

\maketitle

%\begin{figure*}
%\includegraphics{fig_2.ps}% Here is how to import EPS art
%\caption{\label{fig:wide}Use the figure* environment to get a wide
%figure that spans the page in \texttt{twocolumn} formatting.}
%\end{figure*}

%\begin{figure*}
%\centering
%\subfigure[$St_K=0.6, Fr_{p,v}=0.0$]{\label{fig:st1g0P} %% label for first subFig.
%\includegraphics[width=3.5cm]{Fig1a.eps}}\hspace{0.1cm}
%\subfigure[$St_K=0.6, Fr_{p,v}=2.98$]{\label{fig:st1g25P} %% label for first subFig.
%\includegraphics[width=3.5cm]{Fig1b.eps}}\hspace{0.1cm}
%\subfigure[$St_K=5, Fr_{p,v}=0.0$] {\label{fig:st5g0P} %% label for first subFig.
%\includegraphics[width=3.5cm]{Fig1c.eps}}\hspace{0.1cm}
%\subfigure[$St_K=5, Fr_{p,v}=2.98$] {\label{fig:st5g30P} %% label for first subFig.
%\includegraphics[width=3.5cm]{Fig1d.eps}}
%\setlength{\belowcaptionskip}{0pt}
%\setlength{\abovecaptionskip}{0pt}
%\caption{ (Color online) Snapshots of particle positions (red dots) on a vertical x-y slice
%at $z=\pi$. Vorticity contours are superimposed in grayscale. The Reynolds number is $Re_{\lambda}=100.9$.
%The force of gravity is directed downward along the y-axis. There are 7814 particles in each slice.}
%\label{Fig:st1_pc} %% label for entire Fig.
%\end{figure*}

Collisions of heavy particles induced by turbulence are a general and crucial phenomenon with many applications~\cite{Wang_Maxey1993, Wilkinson_Mehlig_Bezuglyy, Falkovich_Punir2007, Gustavsson_Mehlig, Chun2005,  MOKK2012}, including planetesimal formation~\cite{Pan2011}, rain droplet formation~\cite{Bodenschatz_Malinowski_Shaw_Stratmann2010, Devenish_Reeks_Warhaft2012}, and fluidization~\cite{Liu_Gao_Li2005}. Turbulent clustering and relative velocity are two important aspects of particle collision~\cite{Sundaram_Collins1997, Wang_Wexler_Zhou_2000}. Particles suspended in a turbulent flow typically move under gravity~\cite{Csanady1963, Wang_Maxey1993, Yang_Lei1998, Good_Gerashchenko_Warhaft_JFM_2012, Hascoet_Vassilicos2007}. However, the effects of gravity on clustering and the relative motion have seldom been studied until very recently~\cite{Woittiez_Harm_Jonker_Portela2009, Bec_Homann_Ray2014,Gustavsson_Vajedi_Mehlig2014, Park_Lee2014, Parishani_Ayala_Rosa_Wang_Grabowski2015}; thus, an investigation of the effects of settling effects on clustering and the intermittency in the redial relative velocity (RRV)~\cite{Bec_Biferale_Cencini_Lanotte_Toschi2010} is the primary topic of this letter.

The clustering of heavy particles in turbulent or random flows is caused by \textit{centrifugal force}~\cite{Maxey1987,Balkovsky2001,Saw_etal2012a} or \textit{multiplicative amplification}~\cite{Wilkinson_Mehlig_Ostlund_Duncan_2007, Gustavsson_Mehlig2011}, respectively. Particles at a small Stokes number are centrifuged out of regions of high vorticity toward regions of high strain rate and low vorticity~\cite{Maxey1987}. By contrast, clustering  in a random flow at a small Kubo number is determined by the multiplication of many independent random expansion or contraction factors of small volumes spanned by a triad of particle separations~\cite{Gustavsson_Mehlig2011}. If this multiplication decreases after a long time, then the fluctuations in particle number density field will be amplified, and thus clusters can be observed. Here, the Stokes number is $St_K \equiv \tau_{p}/ \tau_{K}$,  where $\tau_{p}$  is the particle relaxation time and $\tau_K$ is the Kolmogorov timescale. The Kubo number is $Ku \equiv u_0\tau_f/ \xi$, where $\xi$ is the smallest characteristic length scale, $\tau_f$ is the smallest characteristic timescale, and $u_0$ is the characteristic velocity scale of the flow~\cite{Duncan_Mehlig_Ostlund_Wilkinson_2005, Wilkinson_Mehlig_2005}.

Previous studies indicate that gravity reduces clustering of particles at small $St_K$~\cite{Falkovich_Fouxon_Stepanov_Nature_2002, Falkovich_Pumir_PoF_2004, Ayala_Rosa_Wang2008}, whereas, very recent results show that gravity may enhance clustering at large $St_K$~\cite{Hascoet_Vassilicos2007, Woittiez_Harm_Jonker_Portela2009, Rosa_njp2013, Bec_Homann_Ray2014, Gustavsson_Vajedi_Mehlig2014, Park_Lee2014}. Gravity-driven enhancement of clustering is attributed to a Gaussian delta-correlated (in time) fluid velocity gradient field or a flow field with a small Kubo number; and the rate of caustic formation is found to be reduced by gravity~\cite{Bec_Homann_Ray2014,Gustavsson_Vajedi_Mehlig2014}. Here, our aims are to study $\tau_g$, the timescale of the Lagrangian correlation of fluid velocity gradient seen by particle pairs under gravity that affects clustering, and furthermore, to explore the mechanism of the remarkable changes in the intermittency in RRV due to gravity.

The direct numerical simulation plus point particle model is typically used to solve for the motion of heavy particles under gravity~\cite{Jin_He_Wang2010, Jin_He_Wang_Zhang2010, Jin_He2013}. The statistical quantities for the flows are listed in Table~\ref{table:table1}. A total of $1.2\times10^6$  particles are tracked. The dynamics of the particles are controlled by the Taylor microscale Reynolds number, $Re_{\lambda}$; $St_K$;  and the Froude number,  $Fr_{g} \equiv a_K/g$, where $a_K$ is the Kolmogorov acceleration and $g$ is the acceleration due to gravity. Because the results for Case $I$ and Case $II$ are similar, we will present the results only for Case $II$.

\begin{table}
\caption{\label{table:table1}  Statistical parameters in isotropic turbulent flows. $N^3$ is the grid number in the flow domain of $({2\pi})^3$. $u'$ is the root mean square (rms) of the velocity. $\varepsilon$ is the energy dissipation rate per unit mass. $L_f$ is the longitudinal integral length scale. $v_K$, $\eta$ and  $\tau_{K} \equiv {(\nu/\varepsilon)}^{1/2}$ are the Kolmogorov velocity scale, length scale and time scales, respectively. $K_{max}=N/3$ is the cutoff wave number in the pseudospectral method.}
\begin{ruledtabular}
\begin{tabular}{ccccccccccccc}
Case& $N^3$  & $Re_{\lambda}$ &$\varepsilon$& $\nu$  &  $u'$ & $L_f$ & $v_{K}$ & $\tau_{K}$ & $\eta k_{max}$ \\
\hline
I&$128^3$& 73.9           &3434.7       &0.095   &19.10  &0.91  &4.31&0.0051&1.10\\
II&$256^3$&100.9           &3468.0       &0.049   &19.52  &0.99  &3.62&0.0037&1.15 \\
\end{tabular}
\end{ruledtabular}
\end{table}

\begin{figure}
\centering
\subfigure{\label{fig:RDF}
\includegraphics[height=6.0cm]{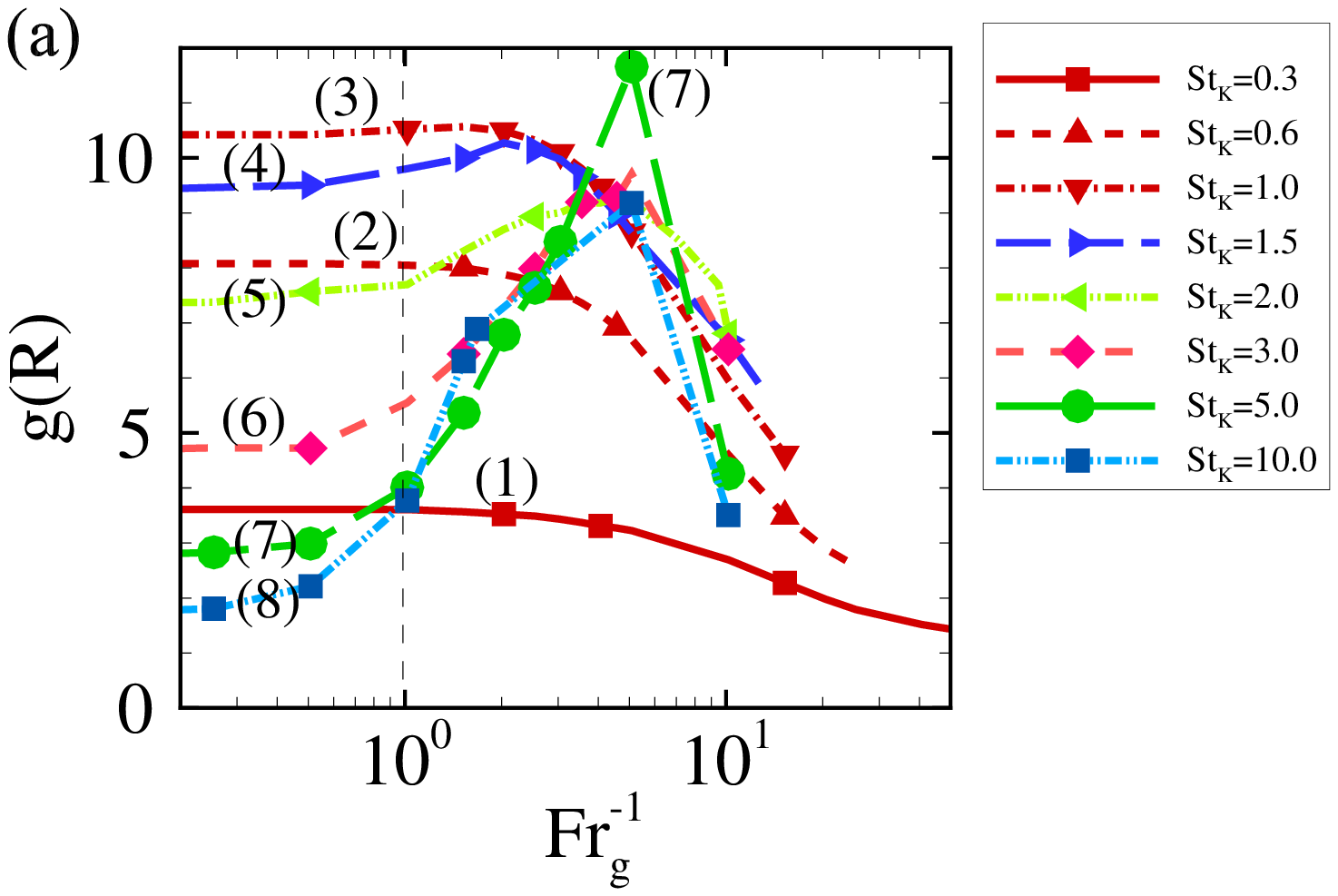}}\hspace{0.2cm}
\subfigure{\label{fig:RDF_Frpa}
\includegraphics[height=6.0cm]{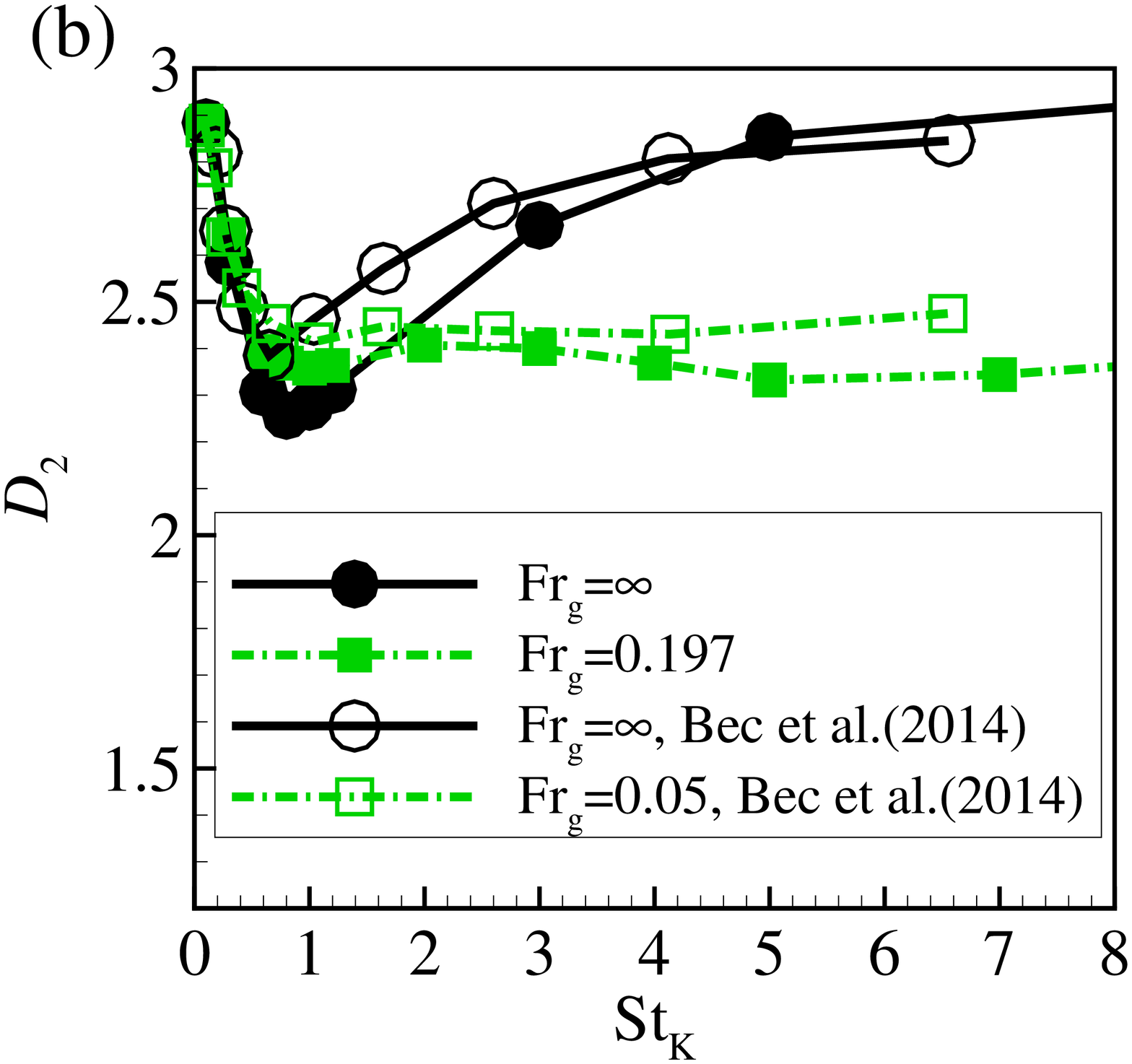}}
\setlength{\belowcaptionskip}{0pt}
\setlength{\abovecaptionskip}{0pt}
\caption{\label{Fig:RDF_PC} (Color online) The degree of clustering as functions of the Froude number and the Stokes number: (a) The RDF, $g(R)$, at a separation $R=0.5\eta$ versus the inverse of the Froude number at different Stokes numbers. (b) The correlation dimension $D_2$ versus the Stokes number at different Froude numbers. The results reported by Bec et al.(2014)~\cite{Bec_Homann_Ray2014} for $Re_{\lambda}=460$, represented by lines with open symbols, are plotted for comparison.}
\end{figure}

Movies showing the effects of gravity on clustering are provided in the Supplemental Material~\cite{SM2},  in which gravity suppresses clustering at $St_K=0.6$, whereas  it enhances clustering at $St_K=5.0$. Similar observations can be found in recent publications~\cite{Bec_Homann_Ray2014, Gustavsson_Vajedi_Mehlig2014, Park_Lee2014}. We use the radial distribution function (RDF), $g(r)$~\cite{Sundaram_Collins1997} and the correlation dimension $D_2$~\cite{Bec_Homann_Ray2014} as measures of clustering. Figure~\ref{Fig:RDF_PC} illustrates $g(R)$ ($R=0.5\eta$) and $D_2$ as functions of the Froude number and Stokes number, respectively. Our results about $D_2$ versus $St_K$ display similar behaviors with those from Ref.~\cite{Bec_Homann_Ray2014}, as shown in Fig.~\ref{fig:RDF_Frpa}. In Fig.~\ref{fig:RDF}, for $St_K\leq1.0$ (Lines~$(1)-(3)$), the RDFs monotonically decrease with an increase in $Fr_{g}^{-1}$. For $St_K>1.0$, (Lines~$(4)-(8)$), the RDFs initially increase to a maximum and then decrease with increasing $Fr_{g}^{-1}$.  Therefore, gravity enhances clustering at $St_K>1$ and reduces clustering at $St_K<1$. The vertical dashed line in Fig.~\ref{fig:RDF} indicates that gravity becomes relevant at $Fr_{g}^{-1}>1$.

\begin{figure}
\begin{center}
\includegraphics[height=6cm]{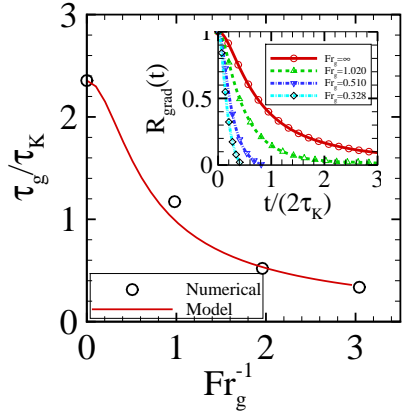}
\caption{(Color online) The timescale of the Lagrangian correlation of the velocity gradient $\tau _{g}$ as a function of $Fr_g^{-1}$ at $St_K=5$. $\tau _{g}$ decreases and it indeed becomes less than $\tau _{K}$ as $Fr_g^{-1}$ increases. The solid line represents a model of the form $\tau _g  = [(1/\tau _{g,0} )^2  + (w_0 /l)^2 ]^{ - 1/2}$.}
\label{Fig:vel_grad}
\end{center}
\end{figure}

The gravity-driven reduction of the clustering at small $St_K$ is due to the ineffectiveness of the centrifugal mechanism because the particles do not have sufficient time to be centrifuged out of regions of high vorticity in the flow when the particles are rapidly going through an eddy under gravity. At large $St_K$,  the reduction in $\tau_g$ due to gravity causes the gravity-driven enhancement of clustering, as demonstrated as follows. The linearized equations for the relative motion of a nearby particle pair, denoted by the relative position $\delta x_{pi}$ and the relative velocity $\delta v_{pi} $, are $ \delta \dot x_{pi}  = \delta v_{pi}$, and  $\delta \dot v_{pi}  = {{[F_{ij} (t)\delta x_{pj}  - \delta v_{pi} ]} \mathord{\left/{\vphantom {{[F_{ij} (t)\delta x_{pj}  - \delta v_{pi} ]} {\tau _p }}} \right. \kern-\nulldelimiterspace} {\tau _p }}$. Here, the dots denote the time derivative $d/dt$; $F_{ij} (t) \equiv {{\partial u_i } \mathord{\left/{\vphantom {{\partial u_i } {\partial x_j }}} \right. \kern-\nulldelimiterspace} {\partial x_j }}({\bf{x}}_p^{(1)}(t), t)$ is the gradient of the fluid velocity experienced by the reference Particle $1$; and $i, j=1,2, 3$, denoting the $3$ direction components. In Fig.~\ref{Fig:vel_grad}, we show how $\tau _g$ varies with $Fr_g^{-1}$ at $St_K=5.0$, where $ \tau _g  \equiv {\rm{ }}\int_0^\infty  {R_{grad} (t)dt}  = \int_0^\infty  {{{\left\langle {F_{ij} (0)F_{mn} (t)} \right\rangle } \mathord{\left/{\vphantom {{\left\langle {F_{ij} (0)F_{mn} (t)} \right\rangle } {\left\langle {F_{ij} (0)F_{mn} (0)} \right\rangle }}} \right. \kern-\nulldelimiterspace} {\left\langle {F_{ij} (0)F_{mn} (0)} \right\rangle }}dt}$ with $i=j=m=n=2$ along the direction of gravity. Note that $F_{22}(t)={{\partial u_2 } \mathord{\left/ {\vphantom {{\partial u_2 } {\partial x_2 }}} \right.
 \kern-\nulldelimiterspace} {\partial x_2 }} =  - ({{\partial u_1 } \mathord{\left/ {\vphantom {{\partial u_1 } {\partial x}}} \right.
 \kern-\nulldelimiterspace} {\partial x}}_1  + {{\partial u_3 } \mathord{\left/ {\vphantom {{\partial u_3 } {\partial x_3 }}} \right.
 \kern-\nulldelimiterspace} {\partial x_3 }})$, denoting the horizontal convergence in an incompressible flow~\cite{Park_Lee2014}.
The Lagrangian correlation $R_{grad}(t)$ versus $Fr_g$ is plotted in the inset. This plot shows that $R_{grad}(t)$ and, thus the timescale $\tau_g$ decay rapidly with decreasing $Fr_g$. We observe that $\tau_g$ can even be less than $\tau_K$ when $Fr_g$ is sufficiently  small. $\tau_g$ can be approximated using a model of the form $\tau_g  = [(1/\tau _{g,0} )^2  + (w_0 /l)^2 ]^{ - 1/2}$, represented by the solid line, where $l$ is the length scale of the fluid velocity gradient, $l=5.6\eta \sim O(\eta)$, and $\tau _{g,0}$ is the timescale of $\tau_g$ when $Fr_{g}=\infty$. $\tau _{g,0}=2.4\tau_K\sim O(\tau_K)$. Thus, we approximate $\tau_g$ using $[(1/\tau _{K} )^2  + (w_0 /\eta)^2 ]^{ - 1/2}= (\eta/w_0)[1+ (v_K/w_0)^2 ]^{ - 1/2} \approx \eta/w_0$ when $ v_K \ll w_0$. Using $\eta$, $v_K$ and $\tau_g$ as the characteristic lengthscale, velocity scale and timescale, respectively, we define an effective Kubo number for a turbulent flow with heavy particles settling within it as $Ku_e \equiv {v_K \tau _g }/\eta = {Fr_g}/{St_K }$. Therefore, $Ku_e\ll 1$ for heavy particles at small $Fr_g$ and large $St_K$ and turbulent flow field seen by settling particles behaves as a random flow with a very short timescale. Multiplicative amplification begins to play a role in clustering enhancement  through many independent random accelerations~\cite{Gustavsson_Vajedi_Mehlig2014}.

\begin{figure}
\centering
\subfigure{\label{fig:PDF_standard_s}
\includegraphics[height=6.0cm]{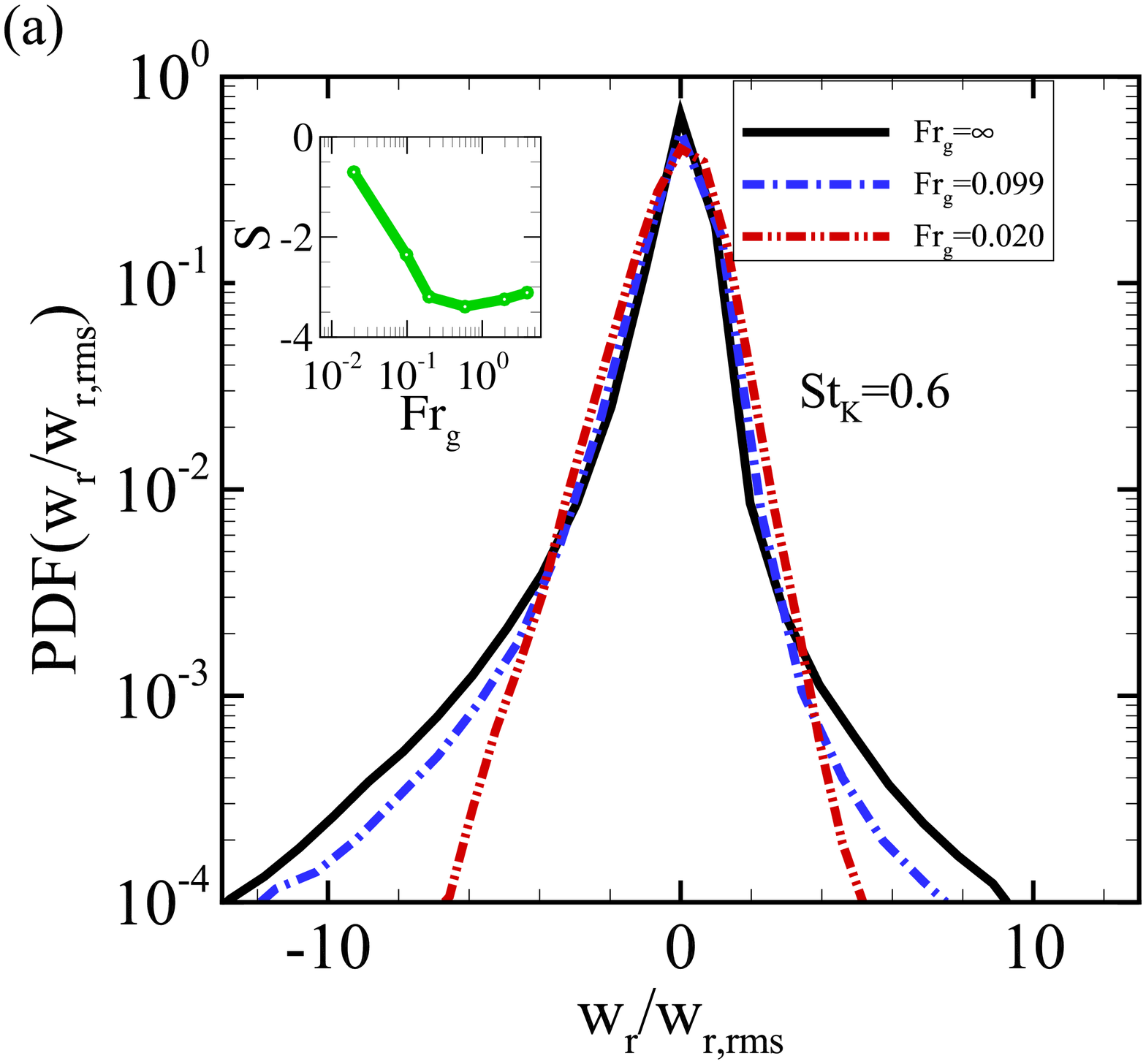}}\hspace{0.2cm}
\subfigure{\label{fig:PDF_standard_l}
\includegraphics[height=6.0cm]{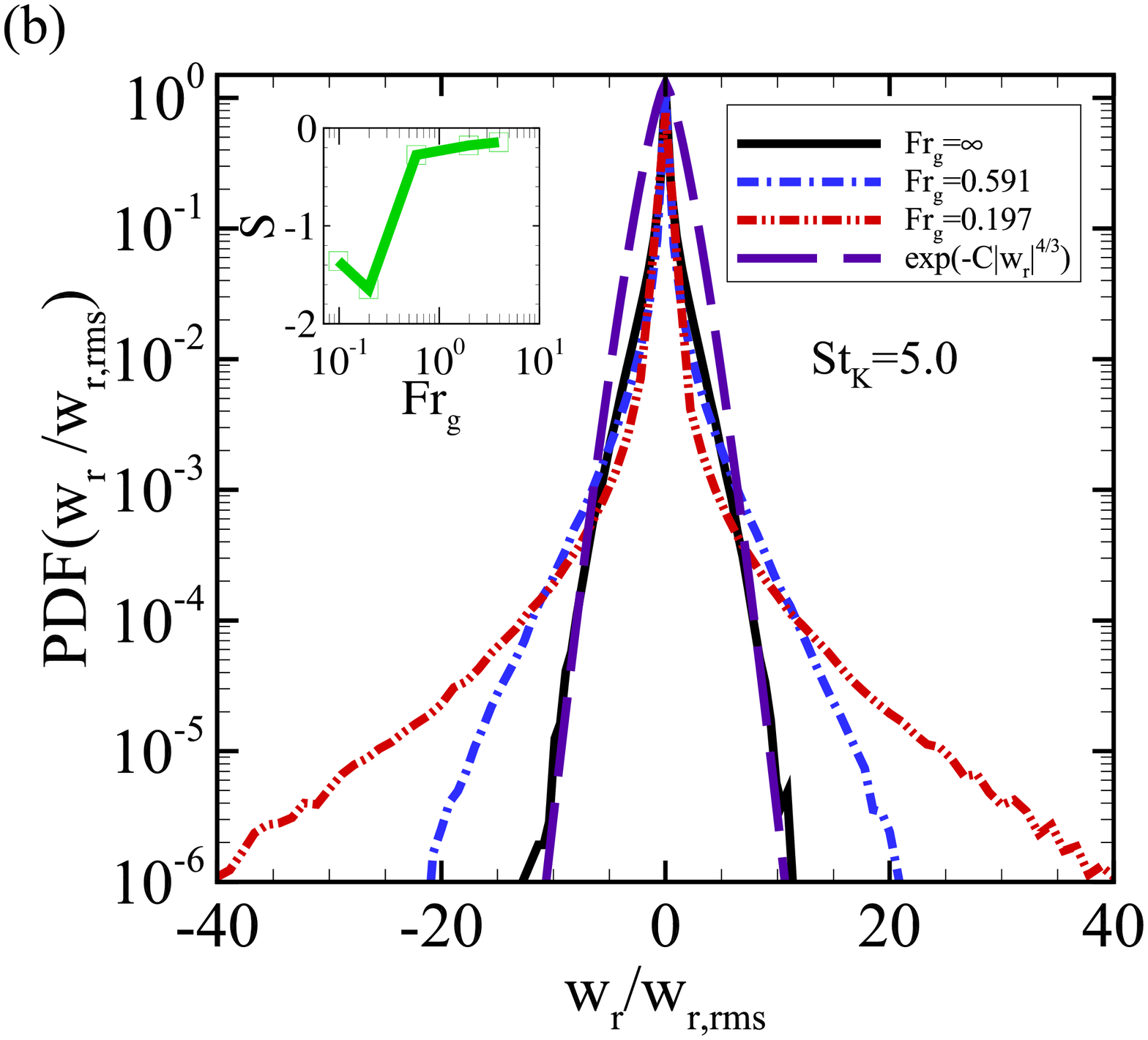}}
\setlength{\belowcaptionskip}{0pt}
\setlength{\abovecaptionskip}{0pt}
\caption{\label{fig:PDF_standard} (Color online) The standardized PDF of the RRVs of heavy particles with a separation $R=0.5 \eta$ versus the Froude number:(a) $St_K=0.6$ and (b) $St_K=5.0$. The insets illustrate the skewness of the PDF as a function of the Froude number. }
\end{figure}

We now turn to the effects of gravity on the statistics of the RRV. For a particle pair consisting of  Particle $1$ and Particle $2$ with a separation $R \equiv \left| {\bf{R}} \right| = | {{\bf{x}}_{p}^{(1)}  - {\bf{x}}_{p}^{(2)} }|$, their RRV, $w_r (R) \equiv ({\bf{v}}_{p}^{(1)} - {\bf{v}}_{p}^{(2)} ) \cdot {\bf{\hat R}}$, is intrinsically related to the clustering. Here, the superscript $(1)$ and $(2)$ denote particles in a pair. ${\bf{\hat R}} = {{\bf{R}} \mathord{\left/
{\vphantom {{\bf{R}} {\left| {\bf{R}} \right|}}} \right. \kern-\nulldelimiterspace} {\left| {\bf{R}} \right|}}$ is the unit vector pointing from Particle $1$ to Particle $2$. When the level of clustering approaches a statistically steady state, the negative skewness, $S = \langle w'_r (R)^3 \rangle /\langle w'_r (R)^2 \rangle ^{3/2}$  denotes the clustering tendency, where $w'_r (R) = w_r (R)/\sqrt {\langle w_r (R)^2 \rangle }$. At $St_K  = 0.6$, as shown in Fig.~\ref{fig:PDF_standard_s}, the PDFs of the RRV are negatively skewed and the absolute value $\left| S \right|$  decreases with decreasing $Fr_g$ as shown in the inset, which means that the degree of clustering also decreases with decreasing $Fr_g$. By contrast, at $St_K = 5.0$, as shown in Fig.~\ref{fig:PDF_standard_l},  $\left| S \right|$ initially increases with decreasing $Fr_g$ down to $Fr_g=0.197$,  implying an enhanced degree of clustering. When $Fr_g$ decreases further,  $\left| S \right|$ begins to decrease, implying that the degree of clustering begins to weaken.  The variations in the skewness at small and large $St_K$ are consistent with the variations in $g(R)$ versus $Fr_g$, as shown in Fig.~\ref{fig:RDF}.

Another interesting observation in Fig.~\ref{fig:PDF_standard} is that the tails of the standardized PDFs of the  RRV exhibit remarkable and opposite changes with decreasing  $Fr_g$ at small and large $St_K$. The tails become increasingly narrower with decreasing $Fr_g$ at $St_K  = 0.6$, whereas they become increasingly broader with decreasing $Fr_g$ at $St_K  = 5.0$. These observations reveal that the intermittency in the RRV is reduced at $St_K=0.6$, but enhanced at $St_K=5.0$ with a decrease in $Fr_g$. Moreover, if gravity is absent, the PDF tail exhibits a shape of the form $\exp [{ - C\left| {w_r } \right|^{4/3} }]$  at a large Stokes number of $St_K  = 5.0$, which is consistent with the prediction of the variable-range projection model proposed for the relative velocities of heavy particles at large Stokes numbers (the long-dashed curve in Fig.~\ref{fig:PDF_standard_l})~\cite{Gustavsson_Mehlig_Wilkinson_Uski_PRL_2008}. We shall interpret this interesting phenomenon from the perspective of the backward-in-time (BIT) relative dispersion.

\begin{figure}
\centering
\subfigure{\label{fig:BIT_S}
\includegraphics[height=6.0cm]{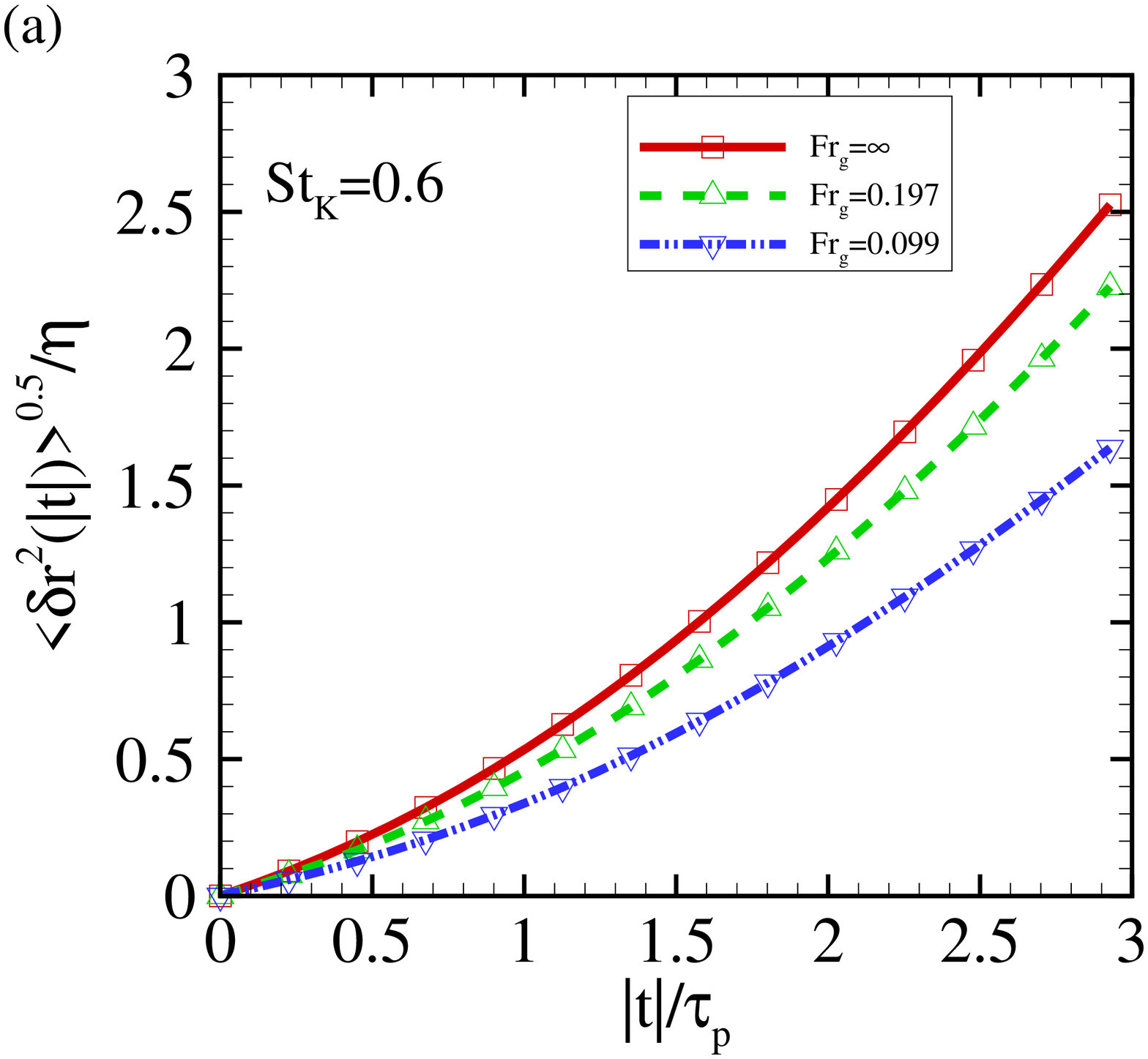}}\hspace{0.2cm}
\subfigure{\label{fig:BIT_L}
\includegraphics[height=6.0cm]{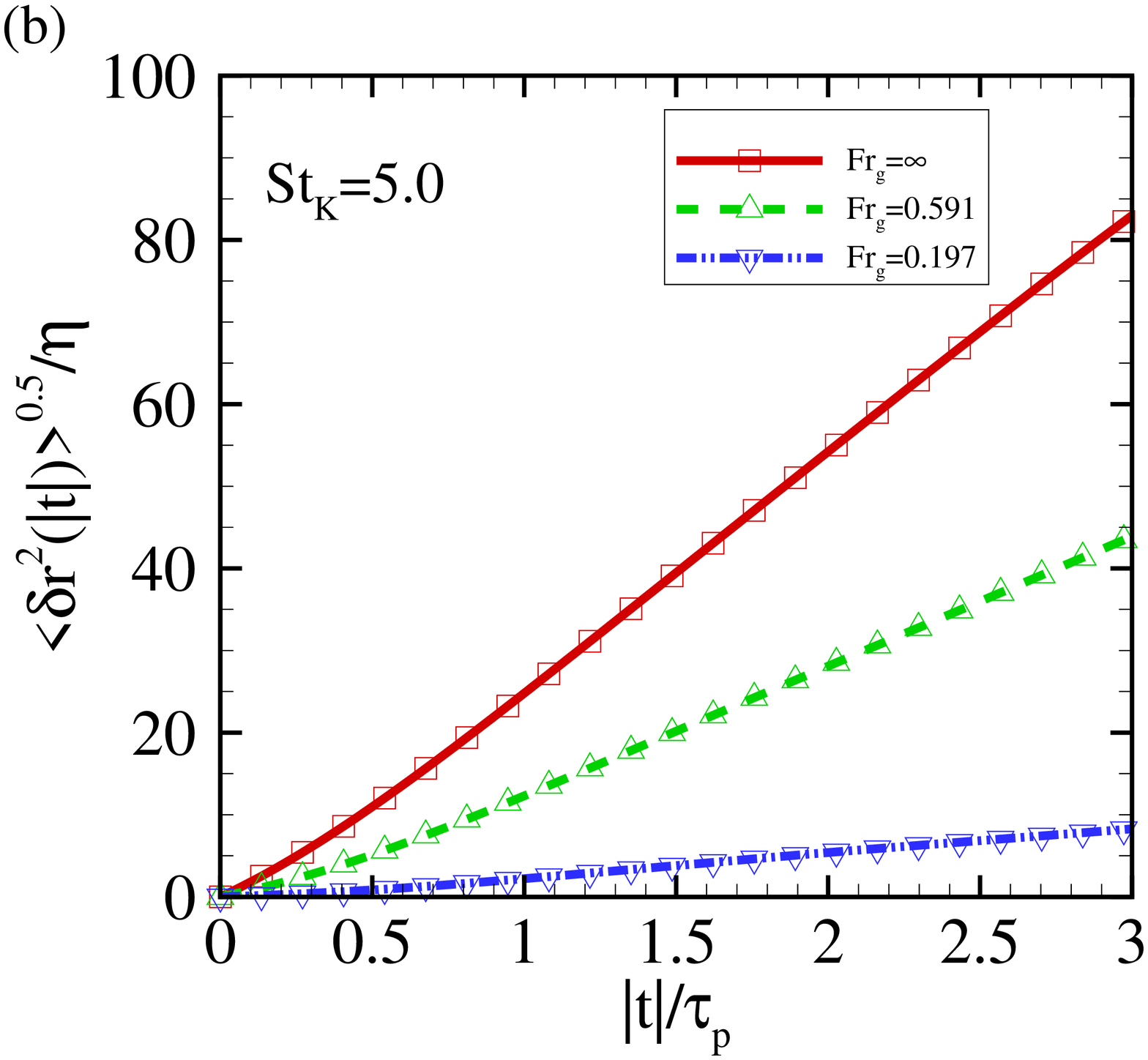}}
\setlength{\belowcaptionskip}{0pt}
\setlength{\abovecaptionskip}{0pt}
\caption{\label{fig:BIT_gravity} (Color online) BIT relative dispersion of heavy particles as a function of Froude number: (a) $St_K=0.6$ and  (b) $St_K=5.0$.}
\end{figure}

The RRV of a particle pair with a separation $R$  can be expressed as  $w_r (R) = \tau _p^{ - 1} {\bf{\hat R}} \cdot \int_{ - 3\tau _p }^0 {\Delta {\bf{u}}({\bf{x}}_{p}^{(1)} (t),{\bf{R}} (t), t){\mathop{\rm exp}\nolimits}({t/ \tau _p }) dt}$,  where $ \Delta {\bf{u}}({\bf{x}}_{p}^{(1)} (t), {\bf{R}}(t),t)$  is the fluid velocity difference for a separation vector ${\bf{R}} (t)$ experienced by the particle pair at $t$, ${\bf{x}}_{p}^{(1)} (t)$  is the position of  the reference particle. Fig.~\ref{fig:BIT_gravity} shows the rms of the separation increment ${\langle \delta r^2 (t)\rangle}^{0.5}  = {\langle |{\bf{R}}(t) - {\bf{R}}(0)|^2 \rangle}^{0.5}$  normalized with respect to $\eta$,  as a function of $Fr_g$ at $St_K  = 0.6$  and $St_K  = 5.0$.  At $St_K  =0.6$  and in the absence of  gravity, ${\langle \delta r^2 (t)\rangle}^{0.5} $  increases within the interval $\left[ { - 3\tau _p ,0} \right]$; however, it remains less than $3\eta$ and  thus it is located in the viscous subrange. Gravity reduces ${\langle \delta r^2 (t)\rangle}^{0.5} $ due to the reduced timescale between particles and eddies. The intermittency of the flow seen by the particle pairs changes very little with a decreasing ${\langle \delta r^2 (t)\rangle}^{0.5}$  because it is approximately saturated at such small separations for a given Reynolds number~\cite{Ishihara_Gotoh_Kaneda2009}. At $Fr_g  = \infty$, particles at such small Stokes numbers are preferentially concentrated in regions of high strain rate with higher intermittency. Gravity pulls these particles out of such highly intermittent regions, and the reference particles ${\bf{x}}_{p}^{(1)}$ become distributed increasingly uniformly throughout the entire field, and the intermittency of the velocity difference seen by the particle pairs decreases, thus, the intermittency of the particle RRV also decreases, as shown in Fig.~\ref{fig:PDF_standard_s}.  By contrast, at $St_K  =5.0$  and in the absence of gravity, ${\langle \delta r^2 (t)\rangle}^{0.5} $ can break free of the viscous subrange and return to the inertial and even the energy-containing subranges within the interval $\left[ { - 3\tau _p ,0} \right]$. It can be as large as  $80\eta$,  such that the Gaussian assumption regarding the particle relative velocity at earlier time used in the variable-range projection model is valid~\cite{Gustavsson_Mehlig_Wilkinson_Uski_PRL_2008}. For particles at large Stokes number, gravitational settling significantly reduces ${\langle \delta r^2 (t)\rangle}^{0.5} $, bringing it from the inertial subrange into the viscous subrange. At $St_K=5.0$, the clustering and ${\bf{x}}_{p}^{(1)}$ are ergodic~\cite{Gustavsson_Mehlig2011}. The intermittency of the flow seen by particle pairs with a small separation ${\langle \delta r^2 (t)\rangle}^{0.5}$  increases, as does the resulting intermittency of the particle RRV, as shown in Fig.~\ref{fig:PDF_standard_l}.The  balance between the mitigation of preferential sampling small-scale flow structure at small $St_K$ and the great reduction in the BIT relative separations at large $St_K$ results in a transition of the variations in the tails of the PDFs occurring at  $St_K \sim3.0$, according to our DNS data.

\begin{figure}
\begin{center}
\includegraphics[height=6cm]{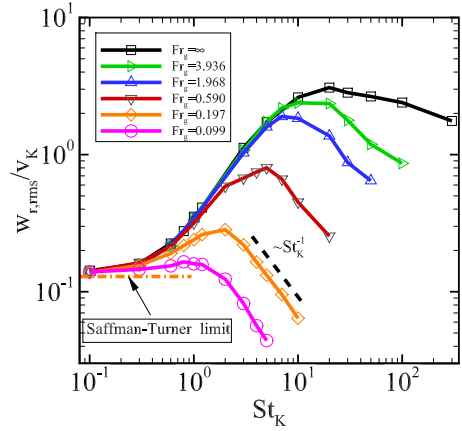}
\caption{(Color online) $w_{r, rms}$ as a function of Stokes number at different Froude numbers. At very small Stokes numbers, the numerical results are consistent with the prediction of the Saffman and Turner limit~\cite{Saffman_Turner_1956}.}
\label{Fig:wr_rms_gravity}
\end{center}
\end{figure}

%\begin{figure}
%\centering
%\subfigure{\label{fig:PDF_non_standard_S}
%\includegraphics[height=6.0cm]{Fig5a.eps}}\hspace{0.2cm}
%\subfigure{\label{fig:PDF_non_standard_L}
%\includegraphics[height=6.0cm]{Fig5b.eps}}
%\setlength{\belowcaptionskip}{0pt}
%\setlength{\abovecaptionskip}{0pt}
%\caption{\label{fig:PDF_non_standard} (Color online) The PDF of the RRV normalized with respect to $v_K$ versus the Froude numbers: (a) %$St_K=0.6$ and (b) $St_K=5.0$. The inset to (b) shows that the PDF of the RRVs of heavy particles can be even narrower that that of the fluid %velocity difference at a Froude number of $Fr_g=0.099$. The shrinkage of the non-standardized PDF tails with decreasing Froude number reduces the %rms of RRV, $w_{r,rms}$, and the rate of caustic formation, as demonstrated in Refs. ~\cite{Bec_Homann_Ray2014, %Gustavsson_Vajedi_Mehlig2014}.}
%\end{figure}
The reduction in BIT relative separation also results in a reduced $w_{r,rms}$.  Fig.~\ref{Fig:wr_rms_gravity} shows  $w_{r,rms}$ as a function of $St_K$ at different $Fr_g$. In the absence of gravity, $w_{r,rms}$  increases with increasing $St_K$, and then slowly decays after reaching a maximum value.  At very large $St_K$ in the absence of gravity, $w_{r,rms} \sim \sqrt {2}u^{'} \sqrt {{{T_{Lp} } \mathord{\left/  {\vphantom {{T_{Lp} } {\tau _p }}} \right. \kern-\nulldelimiterspace} {\tau _p }}}  \propto St_K^{ - 0.5} $~\cite{Abrahamson_1975}, where  $T_{Lp}$ is the timescale of the Lagrangian correlation of fluid velocity experienced by a single-particle.  For a given $Fr_g$, the curve turns downward following a scaling law of the form ${St_K^{-1}}$, rather than  $St_K^{-0.5}$  when settling velocity $w_0=g \tau_p $ becomes large enough. This can be interpreted as follows: for particles at rapid settling velocities, the timescale is $T_{Lp} \sim {L_f \mathord{\left/ {\vphantom {L_f {g\tau _p }}} \right. \kern-\nulldelimiterspace} {g\tau _p }}$. Therefore, $w_{r,rms} \sim {{2u^{'} \sqrt {{L_f/g}}}/\tau_p } \sim St_K^{ - 1}$.

In summary, clustering is monotonically reduced at  $St_K  < 1 $ under gravity because of the ineffectiveness of the centrifugal mechanism, whereas clustering is non-monotonically enhanced at $St_K  > 1 $ because of the multiplicative amplification mechanism in the case that the proposed effective Kubo number is much less than $1$. The mechanism under the opposite variations in the tails of the PDFs of the RRV is explored from the perspective of BIT relative separation and preferential sampling of small-scale flow structures. The rms of RRV is greatly suppressed due to gravity and it follows a scaling law of the form~$St_{K}^{-1}$ at a given $Fr_g$. This understanding of the remarkable changes in the PDF of the RRV challenges  the current intuition that gravity predominantly affects the relative velocity between particles of different sizes falling at different velocities. Therefore, the effects of gravity on clustering and the RRV are a significant influence that must be considered in the parameterization of the geometric collision kernels of monodisperse particles.

\medskip
\begin{acknowledgments}
This work was supported by the National Science Foundation of China (NSFC) (grant numbers 11072247 and 11232011), the National Natural Science Associate Foundation (NSAF) of China (grant number U1230126) and the 973 Program of China (grant number 2013CB834100). GDJ would like to thank Professors L.-P. Wang and H. T. Xu,  Dr. L. B. Pan and Dr. X. Zhao for the helpful discussions during the preparation of this letter. The two anonymous referees are much appreciated for the constructive suggestions to improve the manuscript. GDJ benefitted from the hospitality of the Nordic Institute for Theoretical Physics under the auspices of the program "Dynamics of Particles in Flows: Fundamentals and Applications" in June 2014 in Sweden.
\end{acknowledgments}


\end{document}